\documentclass{ws-procs9x6}

\begin{document}

\title{The gapless 2SC phase \footnote{\uppercase{T}alk 
given by \uppercase{M}ei \uppercase{H}uang.}}

\author{Mei ~Huang, ~ Igor A. Shovkovy}

\address{Institut f\"{u}r Theoretische Physik,\\
       J.W. Goethe-Universit\"{a}t,\\
       D-60054 Frankurt/Main, Germany \\}

\maketitle

\abstracts{Recent results on the gapless 2SC phase are reviewed. These 
include: the thermal stability under the constraint of the local charge 
neutrality condition, the properties at zero and nonzero 
temperatures, and the color screening properties.}

\section{Introduction}

Because the interaction between two quarks in the color anti-triplet 
channel is attractive, sufficiently cold dense quark matter is a color 
superconductor\cite{CS-general}. 
It is very likely that a color 
superconducting phase may exist in cores of compact stars, where bulk 
matter should satisfy the charge neutrality condition as well as 
$\beta$-equilibrium. For a three-flavor quark system, when the strange 
quark mass is small, the color-flavor-locked (CFL) phase \cite{CFL} is 
favorable \cite{N-CFL}. 
For the two-flavor quark system, the charge neutrality condition plays a 
nontrivial role. In the ideal two-flavor color superconducting (2SC) phase, 
the paired 
$u$ and $d$ quarks have the same Fermi momenta. Because $u$ quark carries 
electrical charge $2/3$, and $d$ quark carries electrical charge $-1/3$, 
it is easy to check that quark matter in the ideal 2SC phase is positively 
charged. To satisfy the electrical charge neutrality condition, 
roughly speaking, twice as many $d$ quarks as $u$ quarks are needed. This 
induces a large difference between the Fermi surfaces of the two pairing 
quarks, i.e., $\mu_d - \mu_u = \mu_e \approx \mu/4$, where $\mu,\mu_e$ are 
chemical potentials for quark and electron, respectively. Naively, one would 
expect that the requirement of the charge neutrality condition will destroy 
the $ud$ Cooper pairing in the 2SC phase. 

However, it was found in Ref.~\refcite{N-2SC} 
that a charge neutral two-flavor color superconducting (N2SC) phase does 
exist. Comparing with the ideal 2SC phase, the N2SC phase found in 
Ref.~\refcite{N-2SC} has a largely reduced diquark gap parameter, and the
pairing quarks have different number densities. The latter contradicts
the paring ansatz \cite{enforced}. It is natural to think that the N2SC 
phase found in Ref.~\refcite{N-2SC} is an unstable Sarma state \cite{Sarma}. 
In Ref.~\refcite{g2SC-SH}, it was shown that the N2SC phase is a thermally 
stable state when the local charge neutrality condition is enforced. 
As a by-product, which comes out as a very important feature, it was found 
that the quasi-particle spectrum has zero-energy excitations. Thus, this 
phase was named the gapless 2SC (g2SC) phase. In the following, we 
first show the thermal stability of the g2SC phase, then we discuss its properties 
at zero as well as at nonzero temperatures, at last, we present our most recent 
results about the chromomagnetic instability in the g2SC phase.
 
\section{The gapless 2SC phase and its thermal stability}
\label{stablity} 

Bulk quark matter inside the neutron star should be neutral with 
respect to the color charge as well as the electrical charge. The color 
superconducting phase in full QCD is automatically color neutral 
\cite{colorneutral}. In the Nambu--Jona-Lasinio (NJL) type model, the color
neutrality can be satisfied by tuning the chemical potential $\mu_8$ 
for the color charge. The value of $\mu_8$ for a charge neutral 2SC
phase is very small \cite{N-2SC,g2SC-HS}. Correspondingly, the electrical 
charge neutrality can be satisfied by tuning the chemical potential 
$\mu_{e}$ for the electrical charge. The value of $\mu_e$ is determined 
by the electrical charge neutrality condition. 

The ground state is determined by solving the gap equation together with the
charge neutrality condition. It is found that the ground state of charge neutral
two-flavor quark matter is very sensitive to the diquark coupling constant 
$G_D$ \cite{g2SC-SH}:
\begin{eqnarray}
& G_D/G_S \gtrsim 0.8, & \Delta>\delta\mu,   ~~{\rm 2SC},  \nonumber \\
& 0.7 \lesssim G_D/G_S \lesssim 0.8, & \Delta<\delta\mu, ~~{\rm g2SC}, \nonumber \\
& G_D/G_S \lesssim 0.7, & \Delta=0, ~~{\rm NQM}.
\end{eqnarray}
Where $\delta\mu \equiv \mu_e/2$, $G_S$ is the quark-antiquark coupling constant, 
and ``NQM" indicates the normal phase of quark matter. 
The most interesting case is the g2SC phase, which exists in the diquark coupling
regime $0.7 \lesssim G_D/G_S \lesssim 0.8$. Even though this regime is narrow,
it is worth to mention that, either from the Fierz transformation ($G_D/G_S=0.75$) 
or from fitting the vacuum baryon mass ($G_D/G_S\simeq 2.26/3$) \cite{dnjl2}, the 
value of the ratio $G_D/G_S$ is inside this regime.

The g2SC phase, indicated by the order parameter $\Delta<\delta\mu$, resembles
the unstable Sarma state \cite{Sarma}. For the flavor asymmetric $ud$ quark system,
i.e., when $\mu_e$ is a free parameter and there is no constraint from the charge 
neutrality condition, the solution $\Delta<\delta\mu$ of the gap equation indeed 
corresponds to a maximum of the thermodynamical potential $\Omega_{u,d,e}$. 

However, bulk quark matter inside neutron stars should be charge neutral.  
A nonzero net electrical charge density $n_Q$ will cause an extra energy
$\Omega_{Coulomb} \sim n_Q^2 V^{2/3}$ ($V$ is the volume of the system) 
by the repulsive Coulomb interaction. The total thermodynamical potential
of the whole system is given by $\Omega =  \Omega_{Coulomb} + \Omega_{u,d,e}$. 
The energy density grows with increasing the volume of the system, as a result, 
it is impossible for matter inside stars to remain charged over macroscopic 
distances. So, the proper way to find the ground state of 
the homogeneous neutral $u, d$ quark matter is to minimize the thermodynamical 
potential along the neutrality line $\Omega|_{n_Q=0} = \Omega_{u,d,e}|_{n_Q=0}$
with $\Omega_{Coulomb}|_{n_Q=0}=0$. The g2SC phase corresponds to the global 
minimum of the thermodynamical potential along the charge neutrality line,
thus it is a stable state under the restriction of the charge neutrality condition.

\section{The g2SC phase at zero and nonzero temperatures}
\label{g2SC-T}

As we already mentioned, at zero temperature, in the g2SC phase, the pairing quarks
have different number densities \cite{N-2SC,g2SC-SH}. This is different from the 2SC
phase when $\delta\mu < \Delta$.

It is the quasi-particle spectrum that makes the g2SC phase different 
from the 2SC phase. 
The excitation spectrum for the ideal 2SC phase ($\delta\mu=0$) include: 
two free blue quarks, which do not participate in the Cooper pairing, 
and four quarsi-particle excitations (linear superpositions of 
$u_{r,g}$ and $d_{r,g}$) with an energy gap $\Delta$.  
If there is a small mismatch ($\delta\mu < \Delta$) between the Fermi surfaces 
of the pairing $u$ and $d$ quarks, $\delta \mu$ induces two different branches of 
quasi-particle excitations. One branch moves up with a larger energy gap $\Delta + \delta\mu$,
another branch moves down with a smaller energy gap $\Delta - \delta\mu$.
If the mismatch $\delta\mu$ is larger than the gap parameter $\Delta$, 
the lower dispersion relation for the quasi-particle crosses the zero-energy axis. 
Thus we call the phase with $\Delta < \delta\mu$ the gapless 2SC (g2SC) phase. 
In the g2SC phase, there are only two gapped fermionic quasiparticles, and the 
other four quasiparticles are gapless.

In a superconducting system, when one increases the temperature at a given 
chemical potential, thermal motion will eventually break up the quark 
Cooper pairs. In the weakly interacting Bardeen-Copper-Schrieffer (BCS) 
theory, the transition between the superconducting and normal phases is 
usually of second order. The ratio of the critical temperature 
$T_c^{\rm BCS}$ to the zero temperature value of the gap 
$\Delta_0^{\rm BCS}$ is a universal value\cite{ratio-in-BCS}
$r_{\rm BCS}={T_c^{\rm BCS}}/{\Delta_0^{\rm BCS}} \approx 0.567$.
In the ideal 2SC phase, the ratio of the critical temperature to the 
zero temperature value of the gap is also the same as in the 
BCS theory\cite{PR-sp1}. However, the g2SC phase has very different properties at
nonzero temperatures\cite{g2SC-HS}. The ratio $T_c/\Delta_0$ is not a universal value.
It is infinity when $G_D/G_S \lesssim 0.68$ and approaches $r_{\rm BCS}$
when $G_D/G_S$ increases. The temperature dependence of the gap reveals a 
nonmonotonic behavior. In some cases, the diquark gap could have sizable 
values at finite temperature even if it is exactly zero at zero temperature.
  
\section{Chromomagnetic instability in the g2SC phase}
\label{g2SC-chromo}

Because the g2SC phase has four gapless modes and two gapped modes, one may 
think that the low energy (large distance scale) properties of the g2SC phase 
should interpolate between those of the normal phase and those of the 2SC 
phase. However, its color screening properties do not fit this picture. 

One of the most important properties of an ordinary superconductor
is the Meissner effect, i.e., the superconductor expels the magnetic field.
Using the linear response theory, the induced current 
$j^{ind}_{i}$ is related to the external magnetic field $A_{j}$ as 
$j^{ind}_{i} = \Pi_{ij} A^{j} $, where the response function $\Pi_{ij}$ 
is the magnetic part of the photon polarization tensor. 
The response function has two components, diamagnetic and paramagnetic part. 
In the static and long-wavelength limit, for the normal metal, the paramagnetic 
component cancels exactly the diamagnetic component. In the superconducting 
phase, the paramagnetic component is quenched by the energy gap and produces 
a net diamagnetic response. Thus the ordinary superconductor is a prefect 
diamagnet.
In cold dense quark matter, the gauge bosons connected with the broken 
generators obtain masses in the ideal 2SC phase \cite{Meissner2SC} as 
well as in the CFL phase \cite{Meissner-CFL}, 
which indicate the Meissner screening effect in these phases.

However, in the g2SC phase, it is found that the Meissner screening masses of the five
gluons, corresponding to the five broken generators of the $SU(3)_c$ group,
are ${\it imaginary}$ \cite{Chromo-HS}. This is because, in the static and long-wavelength
limit, the paramagnetic contribution to the magnetic part of these five gluon polarization 
tensors becomes dominant. In condensed matter, this phenomenon is called the paramagnetic 
Meissner effect (PME) \cite{PME}. The imaginary Meissner screening mass indicates a 
chromomagnetic instability of the g2SC phase. There are, several possibilities to 
resolve the instability. One is through a gluon condensate, 
which may not change the structure the g2SC phase. It is also possible that 
the instability drives the homogeneous system to an inhomogeneous phase, like the 
crystalline phase or the vortex lattice phase. This problem remains to be clarified 
in the future. It is also very interesting to know whether 
the chromomagnetic instability develops in the gapless CFL phase\cite{gCFL-AKR}.

\vskip 0.3cm

{\bf Acknowledgements}
The work of M. H. was supported by the Alexander von Humboldt-Foundation, and by the 
NSFC under Grant Nos. 10105005, 10135030. The work of I.A.S. was supported by 
Gesellschaft f\"ur Schwerionenforschung (GSI) and by Bundesministerium f\"ur 
Bildung und Forschung (BMBF).

\end{document}